\documentclass[sigplan,screen]{acmart}

\usepackage{microtype}
\usepackage{wrapfig}
\usepackage{amsmath}
\PassOptionsToPackage{numbers}{natbib}
\usepackage{natbib}
\usepackage{multirow}
\usepackage{booktabs}
\usepackage{hyperref}
\usepackage{diagbox}
\usepackage{multicol}
\hypersetup{
    colorlinks=true,
    linkcolor=blue,
    filecolor=magenta,
    urlcolor=magenta,
}
\usepackage[linesnumbered,algoruled,boxed,lined]{algorithm2e}
\usepackage{algpseudocode}
\usepackage{makecell}
\usepackage[all]{hypcap}
\usepackage{booktabs}
\usepackage{url}
\usepackage{xcolor}
\usepackage{tikz}
\usepackage{diagbox}
\usepackage{listings}
\usepackage{comment} 
\usepackage{graphicx}
\usepackage{longtable}
\usepackage{array}
\usepackage{colortbl}

\theoremstyle{definition}

\definecolor{jsonbackcolour}{rgb}{0.97,0.97,0.96}
\lstdefinestyle{jsonstyle}{
  language=,
  basicstyle=\footnotesize\ttfamily,
  breaklines=true,
  columns=flexible,
  showstringspaces=false,
  frame=lines,
  backgroundcolor=\color{jsonbackcolour},
  literate=
    {:}{{:}}{1}
    {,}{{,}}{1}
    {\{}{{\{}}{1}
    {\}}{{\}}}{1}
    {[}{{[}}{1}
    {]}{{]}}{1}
}

\usepackage{tcolorbox}

\tcbset{
    colback=gray!10, 
    colframe=gray, 
    coltext=black, 
    boxsep=5pt, 
    arc=0mm, 
    top=0mm, bottom=0mm, 
    left=1mm, right=1mm, 
    boxrule=0.5mm, 
    toptitle=1mm, bottomtitle=1mm, 
    titlerule=0mm, 
}
\definecolor{deepgreen}{rgb}{0.0, 0.4, 0.0}

\definecolor{codegreen}{rgb}{0,0.6,0}
\definecolor{codegray}{rgb}{0.5,0.5,0.5}
\definecolor{codepurple}{rgb}{0.58,0,0.82}
\definecolor{backcolour}{rgb}{0.95,0.95,0.92}

\lstdefinestyle{mystyle}{
  backgroundcolor=\color{backcolour}, commentstyle=\color{codegreen},
  keywordstyle=\color{magenta},
  numberstyle=\tiny\color{codegray},
  stringstyle=\color{codepurple},
  basicstyle=\ttfamily\footnotesize,
  breakatwhitespace=false,         
  breaklines=true,                 
  captionpos=b,                    
  keepspaces=true,                 
  numbers=left,                    
  numbersep=5pt,                  
  showspaces=false,                
  showstringspaces=false,
  showtabs=false,                  
  tabsize=2
}
\usepackage{xurl}

\usepackage[framemethod=TikZ]{mdframed}
\definecolor{mycolor}{RGB}{194, 214, 236}

\newcounter{result}

\makeatletter
\g@addto@macro{\@algocf@init}{\SetKwInOut{Parameter}{Parameters}}
\makeatother

\newcommand*\circled[1]{\tikz[baseline=(char.base)]{
              \node[shape=circle, draw, inner sep=0.2pt] (char) {\textcolor{black}{#1}};}}

\lstdefinestyle{yaml}{
     basicstyle=\color{blue}\tiny,
     rulecolor=\color{black},
     string=[s]{'}{'},
     stringstyle=\color{blue},
     comment=[l]{:},
     commentstyle=\color{black},
     morecomment=[l]{-}
 }

\newcommand{\tech}{\mbox{\textsc{FORGE}}} 

\AtBeginDocument{%
  }

\setcopyright{acmlicensed}
\copyrightyear{2026}
\acmYear{2026}
\acmDOI{XXXXXXX.XXXXXXX}
\acmConference[XXX]{XXX}{XXX}{XXX}
\acmISBN{978-X-XXX-XXXX-X/XXX/XX}
 
\begin{document}

\date{}  

\title{Feedback-Driven Execution for LLM-Based Binary Analysis}

\author{XiangRui Zhang}
\affiliation{
  \institution{BeiJing JiaoTong· University}
  \country{China}
}

\author{Qiang Li}
\affiliation{
  \institution{BeiJing JiaoTong· University}
  \country{China}
}

\author{Haining Wang}
\affiliation{
  \institution{Virginia Tech}
  \country{USA}
}

\begin{abstract}

Binary analysis increasingly relies on large language models (LLMs) to perform semantic reasoning over complex program behaviors. 
However, existing approaches largely adopt a one-pass execution paradigm, where reasoning operates over a fixed program representation constructed by static analysis tools. 
This formulation limits the ability to adapt exploration based on intermediate results and makes it difficult to sustain long-horizon, multi-path analysis under constrained context.
We present {\tech}, a system that rethinks LLM-based analysis as a feedback-driven execution process. 
{\tech} interleaves reasoning and tool interaction through a reasoning–action–observation loop, enabling incremental exploration and evidence construction. 
To address the instability of long-horizon reasoning, we introduce a Dynamic Forest of Agents (FoA), a decomposed execution model that dynamically coordinates parallel exploration while bounding per-agent context. 
We evaluate {\tech} on 3,457 real-world firmware binaries. 
{\tech} identifies 1,274 vulnerabilities across 591 unique binaries, achieving 72.3\% precision while covering a broader range of vulnerability types than prior approaches. 
These results demonstrate that structuring LLM-based analysis as a decomposed, feedback-driven execution system enables both scalable reasoning and high-quality outcomes in long-horizon tasks.

\end{abstract}


\maketitle

\section{Introduction}

\sloppy

Binary vulnerability analysis is a representative instance of long-horizon program analysis tasks, especially for closed-source firmware and commercial binaries deployed in real-world systems~\cite{christensen2020decaf, chen2016towards, busch2023teezz, han2023queryx}.  
Despite decades of research, existing approaches largely follow a \emph{one-pass analysis paradigm}: static analysis tools first construct a global program representation, and vulnerability detection operates over this fixed view using rule-based matching or symbolic reasoning~\cite{flawfinder,cwe-checker,redini2020karonte}.  
Recent LLM-based methods~\cite{hussain2025vulbinllm,liu2025llm,chen2025clearagent} also adopt this formulation, where reasoning is performed over precomputed slices of disassembly or decompiled code.
From a systems perspective, this paradigm corresponds to a monolithic execution model, where analysis is performed over a fixed, precomputed representation without feedback from intermediate results.
This limitation is particularly pronounced in binary analysis. 
Unlike source code, binaries lack high-level semantic information such as types, variable names, and structured control flow. 
As a result, analysis must incrementally reconstruct semantics from low-level artifacts (e.g., assembly instructions, indirect calls, and partial dataflows), making it inherently difficult to construct a complete and reliable global representation in a single pass.

\begin{figure}[!t]
\centering
\includegraphics[width=2.8in]{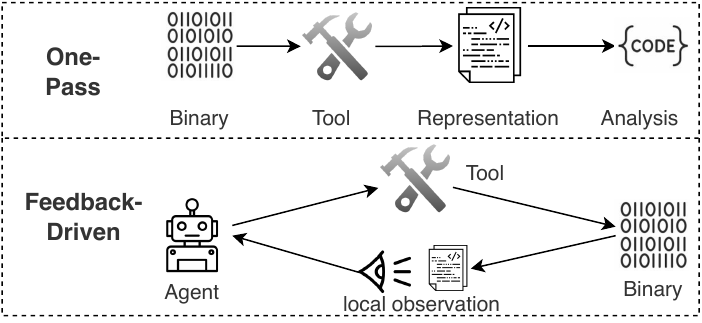}
\caption{Execution model: one-pass vs. feedback-driven. 
}
\label{fig:execution}
\end{figure}

Notably, this formulation introduces a fundamental limitation: \emph{reasoning is decoupled from analysis}.
Once the representation is constructed, exploration strategies cannot adapt based on intermediate results, nor refine hypotheses dynamically.
This limitation is fundamentally an execution issue: the system lacks a mechanism to incorporate runtime feedback into subsequent analysis decisions.
In contrast, real-world binary analysis is inherently feedback-driven.
Expert analysts continuously interleave reasoning with analysis operations, refining their understanding step by step through repeated tool interactions.
Figure~\ref{fig:execution} contrasts the traditional one-pass execution model with a feedback-driven alternative.
This difference is not merely a matter of analysis strategy, but reflects two fundamentally different execution models: one operates over a fixed representation, while the other evolves through iterative interaction with the target binary.
In the feedback-driven model, reasoning, tool invocation, and observation form a closed loop, where intermediate results directly influence subsequent decisions.
Such a mechanism is essential in binary analysis, where semantic information is incomplete and must be incrementally reconstructed from low-level evidence.

Recent advances in Large Language Models (LLMs) make such feedback-driven execution practically realizable~\cite{yang2023intercodestandardizingbenchmarkinginteractive, wu2024oscopilotgeneralistcomputeragents, yang2024swe, xie2024osworldbenchmarkingmultimodalagents}.
By interacting with analysis tools in a loop, LLM agents can incrementally explore binaries, interpret program behavior, and construct execution traces over time.
However, scaling this execution paradigm introduces new challenges.
Binary analysis often requires long sequences of dependent reasoning steps (depth) and simultaneous exploration of multiple candidate paths (breadth).
From a systems perspective, this leads to challenges in managing execution depth, branching factor, and intermediate state under limited context capacity.
Under the constraints of LLMs, these properties lead to context degradation, error accumulation, and instability in reasoning traces.
Thus, the central challenge is to design an execution model that maintains stable and coherent reasoning processes across long-horizon, multi-path analysis.

To address these challenges inherent to binary analysis, we propose {\tech}, an end-to-end LLM-driven system designed to stabilize iterative binary analysis.  
At its core, {\tech} introduces a Dynamic Forest of Agents (FoA), which recursively decomposes analysis tasks and dynamically instantiates agents for parallel exploration.  
FoA adapts its structure according to the evolving analysis state, limiting per-agent reasoning horizons while preserving global exploration coverage.  
This design transforms analysis from a monolithic execution into a decomposed and feedback-driven system.
It prevents long reasoning chains from collapsing and mitigates context drift across divergent paths, enabling stable reasoning over complex binaries.

Building on this stabilized execution model, {\tech} further enables a unified discovery–verification workflow.  
Intermediate results—such as call-chain fragments, symbolic values, and taint flows—are continuously recorded during analysis, forming structured evidence chains.  
These chains can be directly reused in a verification stage, allowing vulnerability discovery and validation to be performed within the same framework.

We evaluate {\tech} on 3,457 real-world firmware binaries and compare it with state-of-the-art approaches, including Mango~\cite{gibbs2024operation}, SaTC~\cite{chen2021sharing}, LATTE~\cite{liu2025llm}, and SWE~\cite{jimenez2023swe, yang2024swe}.  
{\tech} identifies 1,274 vulnerabilities across 591 unique binaries, achieving 72.3\% precision and covering a broader range of vulnerability types.
These results demonstrate that stabilizing iterative analysis leads to both improved detection performance and more consistent vulnerability discovery across complex binaries.
They further suggest that controlling execution structure is a first-order factor in scaling LLM-based analysis systems.

Our main contributions are summarized as follows:
\begin{itemize}
\item \textit{Execution formulation.} 
We reformulate LLM-based analysis as a feedback-driven execution problem under partial observability, highlighting the limitations of one-pass reasoning paradigms.
\item \textit{Decomposed execution model.} 
We propose FoA, a dynamic execution model that stabilizes long-horizon reasoning through recursive decomposition and parallel exploration under a bounded context.
\item \textit{Execution–validation integration.} 
We show how structured execution naturally produces replayable evidence, as a unified discovery–verification framework.
\item \textit{End-to-end system and evaluation.} 
We implement {\tech}~\footnote{Code and Artifacts are available at \url{https://github.com/bjtu-SecurityLab/FORGE}}  and demonstrate its effectiveness and scalability on large-scale real-world experiments.
\end{itemize}

\noindent
\textbf{Roadmap}. 
Section~\ref{sec:background} reviews the background and technical challenges. 
Section~\ref{sec:design} details task decomposition, agent generation, and coordination mechanisms. 
Section~\ref{sec:discovery-validation} presents how {\tech} unifies vulnerability discovery and validation under the same FoA architecture. 
Section~\ref{sec:exp} describes the experimental results. 
Section~\ref{sec:discuss} discusses the limitations, and Section~\ref{sec:related} reviews related work. 
Finally, Section~\ref{sec:conclusion} concludes the paper.

\fussy 
\section{Background}
\label{sec:background}

\subsection{Binary Vulnerability Detection: One-Pass vs. Iterative Paradigms}

Binary vulnerability detection has traditionally been formulated as a \emph{one-pass analysis problem}, which can be viewed as a static computation model.  
In this paradigm, static analysis tools—such as disassemblers, control-flow graph (CFG) builders, and symbolic execution engines—are executed once to construct a global program representation, over which vulnerability detection is performed.  
This formulation assumes that a sufficiently complete representation can be constructed upfront, and that reasoning can be applied independently of the analysis process itself.  
This abstraction underlies a wide range of existing approaches, including pattern-based tools (e.g., Flawfinder~\cite{flawfinder}, CWE-checker~\cite{cwe-checker}) and symbolic execution systems (e.g., Karonte~\cite{redini2020karonte}).  
Recent LLM-based methods~\cite{hussain2025vulbinllm,liu2025llm,chen2025clearagent} also follow this formulation by performing reasoning over precomputed code representations.  

In contrast, binary analysis can be more accurately modeled as a \emph{sequential decision process under partial observability}.  
Rather than operating on a fixed representation, the analysis process must iteratively select actions (e.g., disassembling functions, resolving dependencies), observe partial program state, and update its reasoning accordingly.  
Each decision influences subsequent exploration, and the program representation itself is progressively constructed during execution.  

This distinction fundamentally changes the nature of the problem: analysis is no longer a one-pass computation over a static structure, but an adaptive process that interleaves reasoning and analysis under incomplete information.  
This shift introduces new system-level requirements, as the execution process must support sequential decision-making, maintain intermediate state across steps, and dynamically adapt exploration based on partial observations.

\subsection{LLM-Driven Iterative Analysis}

An execution model aligned with the iterative paradigm can be realized through LLM-based agents.
An LLM agent forms a \emph{reasoning–action–observation loop}.  
At each step, the agent selects an analysis action (e.g., disassembling a function, tracing a call), receives a localized program fragment, and updates its reasoning before deciding the next action.  
This process may involve hundreds of iterations, gradually constructing an understanding of the binary.

Within this iterative paradigm, reasoning and static analysis are tightly interleaved, leading to two key advantages.
\textit{(1) Semantic-aware reasoning.}  
By incrementally analyzing relevant code regions and refining hypotheses, the agent can interpret program behavior beyond fixed patterns.  
It captures implicit relationships (e.g., sanitization logic and control dependencies) and dynamically adapts its analysis strategy.
\textit{(2) Evidence-preserving analysis.}  
Each step in the reasoning–action loop produces concrete artifacts, including disassembly snippets, addresses, call chains, and intermediate states.  
These artifacts form structured evidence chains that can be replayed to validate detected vulnerabilities.

\subsection{Challenges in Scaling Iterative Analysis}

Unlike one-pass approaches, the iterative execution model introduces distinct structural properties.
(1) Partial observability.  
The agent only observes a small portion of the program at each step and must decide what to explore next without a global view.
(2) Sequential dependency.  
Each decision influences subsequent exploration, making early reasoning steps critical to the overall analysis outcome.
(3) Trace-based execution.  
The analysis process produces a sequence of reasoning steps and tool invocations, forming an explicit execution trace that records how conclusions are derived.
These properties, while enabling flexible and adaptive analysis, also introduce fundamental challenges when scaling to large binaries.

\begin{figure}[!t]
\centering
\includegraphics[width=3.1in]{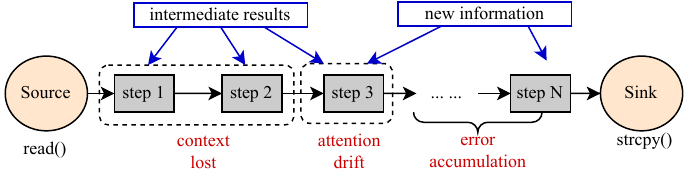}
\caption{Reasoning Depth: Long analysis chains suffer from context degradation, error accumulation, and attention drift.}
\label{fig:longchain}
\end{figure}

\textit{Reasoning depth.}  
Even moderately sized binaries can contain millions of instructions with complex control and data dependencies.
As the analysis depth increases, earlier information is easily lost, errors accumulate, and the system may lose focus on the original taint source.
Single-agent methods (e.g., ReAct~\cite{yao2022react}) degrade rapidly beyond a handful of steps, while multi-agent frameworks with static workflows (e.g., AutoGen~\cite{wu2023autogen}) cannot dynamically adjust to variable-length reasoning chains.
Figure~\ref{fig:longchain} illustrates this degradation.

\begin{figure}[!t]
\centering
\includegraphics[width=3.0in]{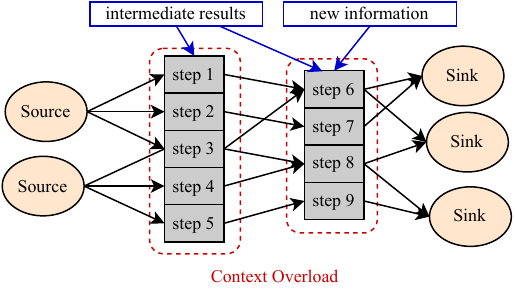}
\caption{Reasoning Breadth: Simultaneous analysis of multiple source-sink combinations leads to context overload and selection bias.}
\label{fig:breadth}
\end{figure}

\textit{Reasoning Breadth.}
Large binaries often contain multiple sources, sinks, and propagation paths that must be analyzed concurrently.
Tracking all relevant paths without omissions or confusion is challenging for a single LLM.
Static multi-agent frameworks lack dynamic coordination, making simultaneous path analysis prone to overload or selection bias.
Figure~\ref{fig:breadth} illustrates this combinatorial challenge.

The central challenge is therefore not enabling semantic reasoning or evidence generation—both are naturally supported in iterative analysis—but maintaining \emph{stable and coherent execution processes} across long-horizon and multi-path analysis.
This requires a system design that supports dynamic exploration while preserving coherence across both depth and breadth.

\section{System Design}
\label{sec:design}

This section presents the design of {\tech} as an execution model for long-horizon binary analysis.
{\tech} is built upon the \textit{Forest of Agents (FoA)} model, which organizes analysis as a dynamically expanding set of interacting agents. 

\begin{figure*}[!t]
\centering
\includegraphics[width=5.3in]{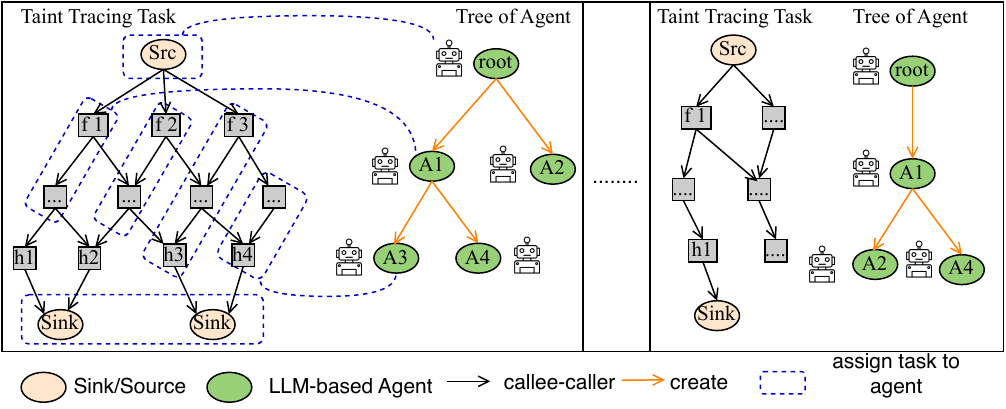}
\caption{FoA as a dynamic execution model. Each tree corresponds to a source-rooted analysis process. 
Nodes represent agents, and edges denote task decomposition and propagation.}
\label{fig:foa}
\end{figure*}

\subsection{FoA Execution Model}

FoA models binary analysis as a dynamically constructed computation structure in the form of a forest of agent-executed task trees:
\[
FoA = \{\mathcal{T}_1, \mathcal{T}_2, ..., \mathcal{T}_n\},
\]
where each tree $\mathcal{T}$ corresponds to an analysis process rooted at a specific source.
Each node in a tree is an agent $A_i(T_i)$, where $T_i$ is an exploration task assigned to the agent.
An agent represents the minimal unit of reasoning and execution in {\tech}, encapsulating a localized analysis task over a bounded program context.
Agents are strictly organized in parent–child relationships, ensuring that the FoA maintains a well-defined hierarchical structure.
Collectively, these trees form a dynamically evolving execution FoA.

\textbf{FoA as a Runtime Structure.}
FoA is not constructed upfront; instead, it is incrementally materialized as the analysis proceeds. 
At any point during execution, only a subset of the full exploration structure is instantiated, corresponding to the portions of the analysis that have been explored.
This implies that FoA is inherently partial: 
unexplored branches remain implicit, and only those required for resolving the analysis objective are realized at runtime.
As a result, FoA can be viewed as a partially materialized computation graph whose shape is dynamically determined during execution.
As such, the FoA represents not only the exploration structure but also the evolving execution state of the analysis.

As illustrated in Figure~\ref{fig:foa}, FoA organizes agents to collaboratively accomplish binary vulnerability detection.
Each tree represents a taint tracing task for a source, while agents perform semantic reasoning, produce intermediate evidence, and make path pruning decisions.
This process naturally forms a branching structure that captures alternative execution paths and data dependencies.
Different branches may stop expanding at different depths depending on local analysis decisions, resulting in an irregular, demand-driven structure.

\textbf{Execution Semantics.}
FoA execution can be understood as the composition of two tightly coupled processes over this dynamically constructed structure.
\textit{(1) Forward (structure materialization).}
The forward process incrementally constructs the FoA by expanding agents into sub-tasks.
It defines the exploration space of the analysis, determining which program paths and data flows are examined.
\textit{(2) Backward (evidence propagation).}
In parallel, the backward process propagates analysis results along the constructed structure.
Each agent continuously produces structured evidence fragments during its reasoning process, including code locations, taint states, and semantic interpretations.
These fragments are returned to parent agents and recursively aggregated.
A complete evidence chain emerges when evidence from descendant agents is successively integrated along a path to the root, yielding a coherent explanation of a vulnerability.

Together, these two processes define the execution semantics of FoA: 
the forward process determines the reachable computation space, while the backward process defines how partial results are composed into semantically valid outcomes.

\subsection{Task Decomposition and Agent Generation}

FoA constructs its analysis through a forward execution process, which progressively expands from a source task into a hierarchy of subtasks and agents. 
Rather than predefining a fixed execution graph, the system dynamically builds an exploration structure at runtime.
Starting from an initial task $T_0$, FoA performs recursive decomposition and delegation:
\[
\begingroup
\setlength{\arraycolsep}{3pt}
\begin{array}{@{}c c c c c c c@{}}
T_0 & \;\rightarrow\; & \{T_1, T_2, \dots\} & \;\rightarrow\; & \cdots & \;\rightarrow\; & \{T_{i}^{(j)}\} \\[-2pt]
\downarrow & \phantom{\;\rightarrow\;} & \downarrow & \phantom{\;\rightarrow\;} & \downarrow & \phantom{\;\rightarrow\;} & \downarrow \\[-2pt]
A_0 & \;\rightarrow\; & \{A_1, A_2, \dots\} & \;\rightarrow\; & \cdots & \;\rightarrow\; & \{A_{i}^{(j)}\}
\end{array}
\endgroup
\]
The upper row captures recursive task decomposition, 
while the lower row represents the corresponding runtime agent structure. 
Each task $T_i$ is not executed directly; instead, it is realized through a delegation operation that instantiates an agent $A_i$.
Formally, delegation is performed via a tool-mediated operation:
\[
A_j = \text{Delegate}(T_j)
\]
where the delegation operator encapsulates agent creation and task passing. 
Thus, FoA execution alternates between two tightly coupled steps:
(1) task decomposition, which produces new tasks, and 
(2) delegation, which maps tasks to runtime agents.

\textbf{Task Decomposition.}
An exploration task is the unit of recursive taint tracing. We represent a task as
\[
T(f, e, s, o)
\]
where $f$ is the current function (with its disassembly address); $e$ is the taint entry (the register/stack slot/argument in $f$ that receives the taint); $s$ is the taint source (the origin expression or call that produced the taint); $o$ is the analysis objective (a concise, machine-readable description of what to prove about the propagation, e.g., ``can value in $e$ reach an argument of \texttt{system} without sanitization'').
An exploration task is recursively decomposed into a set of subtasks:
\[
T(f, e, s, o) \rightarrow \{T(f_1, e_1, s_1, o_1), \ldots, T(f_n, e_n, s_n, o_n)\}
\]
where each child task corresponds to a refined analysis subproblem derived from the current task context (e.g., following a data dependency or exploring a control-flow branch).
The decomposition is determined by the agent's reasoning process, which is instantiated using an LLM but constrained by the task structure.
Subtasks may represent sequential steps, parallel branches, or alternative exploration paths.

\begin{figure}[!t]
\centering
 \includegraphics[width=.92\linewidth]{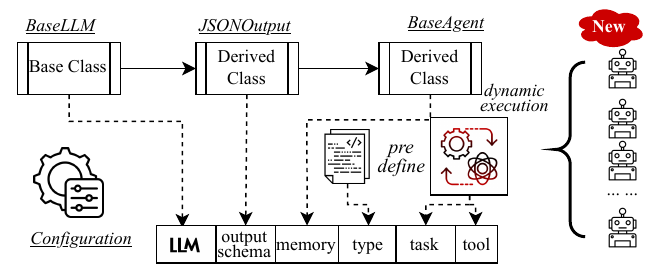}
\caption{Dynamic Agent Generation Process}
\label{fig:generation}
\end{figure}

\textbf{Delegation as Agent Generation.}
FoA treats delegation as a first-class runtime primitive that defines how computation is dynamically expanded and distributed.
Each subtask $T_i$ is passed as input to a delegation operator, which instantiates a new agent: $T_i\xrightarrow{\text{delegate}}A_i$.
One possible instantiation of this abstraction is to realize delegation through a tool-mediated interface, which triggers the creation of a new agent responsible for executing the task.
As illustrated in Figure~\ref{fig:generation}, AgentTool acts as a bridge between agent reasoning and runtime instantiation: 
subtasks are first produced by the agent's reasoning process, and then materialized as new agents via tool invocation. 
Each agent is equipped with AgentTool, enabling it to create new agents as needed. 
The generated agent $A_i$ is responsible for executing $T_i$ independently, including invoking tools, interacting with the environment, and producing intermediate results. 
This abstraction elevates delegation from an implementation detail to a first-class runtime primitive, defining how computation is dynamically distributed across agents in the FoA.

For example, a concrete implementation (AgentTool) may support both parallel and sequential delegation strategies:
\circled{1} \textit{Parallel Generation (1$\rightarrow$n)}. When the LLM identifies multiple callee functions requiring exploration, the current agent creates multiple child agents, each responsible for one distinct subtask $(f_i, e_i, s_i, o_i)$. These child agents can execute concurrently, enabling true runtime parallelism and efficient handling of reasoning breadth across multiple paths.
\circled{2} \textit{Sequential Generation (1$\rightarrow$1)}. When the LLM identifies a single critical callee requiring deep tracing, the agent creates one child agent to continue exploration along that specific path. This ensures that long reasoning chains are preserved without overloading any single agent.


\subsection{Agent Execution and Evidence Formation}

\textbf{Agent Execution.}
{\tech} models agent execution as a constrained execution interface between the reasoning process and the target binary.
In this design, the binary serves as the external environment, and tools provide the only observable channel through which agents can query program state and behavior.
Each agent step consists of invoking a tool with structured parameters derived from its current task context (e.g., querying disassembly, control-flow relations, or data dependencies), and incorporating the returned results into its reasoning process. 
Thus, tool invocations define the operational semantics of agent–environment interaction, while remaining orthogonal to the FoA structure itself.

\textbf{Evidence Formation.}
Rather than exposing raw tool outputs, agents produce structured evidence as execution artifacts.
These artifacts constitute a persistent representation of the execution trace, enabling replay and verification.
An evidence object summarizes the essential information required to describe a propagation step, including:
code location (e.g., function address), relevant snippet or instruction context, taint-related variables or registers, and a concise semantic interpretation of the propagation. For example:
\begin{quote}
\texttt{\{"next\_function": "sub\_401B20", "evidence": \{"addr": 0x401B20, "snippet": "...", "vars": {...}, "note": "propagated via R0"\}\}}
\end{quote}
This abstraction decouples low-level tool outputs from higher-level reasoning, enabling consistent aggregation and interpretation across independently executed agents.

This evidence serves two purposes:
(1) it compresses low-level tool outputs into semantically meaningful units, and
(2) it provides a uniform interface for cross-agent communication and aggregation.

\subsection{Runtime Coordination and Aggregation}

FoA execution is governed by three tightly coupled runtime mechanisms: 
(1) forward expansion constraints, 
(2) hierarchical aggregation, and 
(3) decentralized coordination. 
Together, these mechanisms ensure that the dynamically constructed FoA remains bounded, consistent, and semantically coherent.

\textbf{Forward Expansion Constraints.}
FoA expansion is not centrally scheduled; instead, it is driven by local agent decisions under structural constraints.
First, expansion preserves the hierarchical structure defined by FoA: tasks can only be delegated along parent–child relationships.
New agents can only be generated through parent–child delegation, i.e., $T_i$ can only produce subtasks $\{T_j\}$ that are delegated to child agents $\{A_j\}$. 
Second, each agent operates within a strictly bounded scope defined by its task $T(f,e,s,o)$. 
The parent agent passes only task-relevant context to the child, including the target function, taint entry, and analysis objective. 
This ensures that reasoning remains localized and avoids uncontrolled context growth.
Third, expansion decisions are fully delegated to the LLM. 
At each step, an agent determines whether to:
(i) generate new subtasks for further exploration, or 
(ii) stop expansion for the current path.

Termination is therefore not a separate phase, but a natural outcome of the expansion process. 
An agent stops generating new tasks when the LLM determines that the current path is either sufficiently resolved (e.g., reaching a vulnerability-relevant endpoint) or no longer worth exploring. 
This design enables adaptive, demand-driven exploration while preventing unbounded growth.

\textbf{Hierarchical Aggregation Model.}
While forward expansion constructs the FoA, execution results are propagated in the reverse direction through hierarchical aggregation.
Each agent produces a structured result upon completion of its assigned task. 
This result is not merely a tool return; instead, it is a runtime-level output that encapsulates:
(i) the local analysis outcome, and 
(ii) a structured evidence fragment derived from tool interactions.

When a child agent finishes execution, its result is returned to the parent agent as part of the runtime control flow (rather than as a direct tool response). 
The parent agent incorporates this result into its local reasoning state, potentially combining multiple child results.
This recursive return-and-integration process forms a hierarchical aggregation structure: leaf agents produce atomic evidence fragments, which are progressively merged along the tree edges. 
A complete evidence chain is formed when results from leaf agents are successively integrated up to the root, yielding a coherent explanation of the full propagation path.

Importantly, aggregation is tightly coupled with execution: a parent agent may suspend its own reasoning until child results are available, making aggregation an inherent part of the runtime rather than a post-processing step.
Thus, aggregation defines how distributed execution results are composed, forming a core part of the system’s execution semantics rather than a passive collection mechanism.

\textbf{Decentralized Coordination.}
FoA does not rely on a centralized planner or global scheduler; instead, execution control is fully embedded in the runtime structure itself.
Instead, coordination emerges from local interactions between agents.
Each agent independently decides how to proceed based on its local context and received evidence. 
Parent agents control the lifecycle of their children through delegation and aggregation, while child agents operate independently within their assigned scope.

This decentralized design enables:
(i) adaptive exploration across multiple paths, 
(ii) parallel execution of independent branches, and 
(iii) robustness to partial failures or inconsistent intermediate results.
Overall, FoA enforces a key execution invariant: reasoning is localized, while execution is globally compositional through hierarchical structure.

\section{FoA Instantiation: Vulnerability Discovery and Validation}
\label{sec:discovery-validation}

\begin{figure}[!t]
\centering
\includegraphics[width=3.1in]{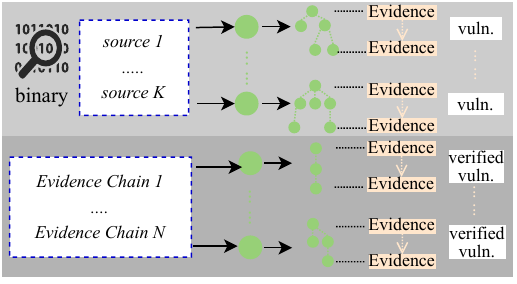}
\caption{FoA for Vulnerability Discovery and Validation.}
\label{fig:discovery-validation}
\end{figure}

We demonstrate how FoA can be instantiated to support a multi-phase vulnerability analysis workflow, consisting of discovery and validation.
While these phases serve distinct purposes, they are both realized as executions of the same FoA model, operating over shared task abstractions and evidence structures.
Figure~\ref{fig:discovery-validation} illustrates a closed-loop execution process:
\textbf{discovery → vulnerability hypotheses (evidence chains) → validation → verified vulnerabilities (evidence chains)}.
These phases correspond to different execution modes over the same FoA structure.
The two modes share identical execution semantics---including task decomposition, delegation, and hierarchical aggregation---and differ primarily in their inputs and task specifications.
Discovery operates over source-initialized tasks to explore the program space, while validation operates over previously generated evidence chains to perform constrained re-execution.
An illustrative 5-turn example of this process is provided in Listings~\ref{appendix:foa_code_example-1}–\ref{appendix:foa_code_example-5} in the appendix.

\subsection{Vulnerability Discovery}

In the discovery phase, root agents are initialized from a set of high-risk sources obtained via root enumeration of the binary. 
Each source serves as the root of a task-specific exploration tree. 
Each agent operates under a discovery-oriented task specification (Listing~\ref{lst:discovery-prompt}), which prioritizes:
\begin{itemize}
\item Evidence-based analysis: All findings must be grounded in concrete evidence extracted via the tools' outputs.
\item Taint identification: Agents identify and validate all externally controllable taint entries.
\item Delegated tracing: New agents perform the actual taint propagation.
\item Focused task execution: Each agent works strictly within its assigned task to ensure comprehensive coverage of exploitable paths.
\end{itemize}

The output of the discovery phase is not a raw list of function names or warnings, but a structured \emph{evidence chain} that records concrete propagation steps and semantic reasoning results. 
Each chain records concrete taint propagation steps, call-chain reasoning traces, and semantic explanations of exploitability.
An example evidence chain is shown in Listing~\ref{lst:discovery-evidence}.
Thus, discovery produces both candidate vulnerabilities and the structured evidence artifacts required for subsequent execution in the validation phase.

\begin{table*}[!t] \small
\centering
\caption{Vulnerability Discovery Comparison between \textit{Operation Mango} and {\tech} on the Karonte Dataset. For \textit{Operation Mango}, Underlying Vulns = CI + BoF; for {\tech}, Underlying Vulns = CI + BoF + Other Vulns.}
\label{tab:performance}
\begin{tabular}{l|cccc|ccccc}
	\toprule
 & \multicolumn{4}{c|}{\textit{Operation Mango}}  & \multicolumn{5}{c}{{\tech}}  \\
	& \makecell[c]{Underlying \\ Vulns} & \makecell[c]{Affected \\ Binaries} & CI Vulns & BoF Vulns & \makecell[c]{Underlying \\ Vulns} & \makecell[c]{Affected \\ Binaries} & CI Vulns & BoF Vulns & Other Vulns \\
\midrule
Netgear & 1447 & 173 & 339 & 1108 & 880 & 429 & 457 & 232 & 191 \\
Tenda   & 391  & 21  & 154 & 237  & 185 & 61  & 100 & 42  & 43 \\
D-link  & 85   & 36  & 8   & 77   & 159 & 74  & 85  &30  & 44 \\
TP-link & 94   & 14  & 9   & 85   & 50  & 27  & 25  & 25  & 0 \\
\textbf{Total} & 2017 & 244 & 510 & 1507 & \textbf{1274} & 591 & 667 & 329 & 278 \\
\bottomrule 
\end{tabular}
\end{table*}

\subsection{Vulnerability Validation}

Unlike discovery, which explores from identified sources, validation is formulated as an evidence-constrained execution process.
It begins with a structured evidence chain generated during discovery and re-executes it within the FoA framework by verifying each propagation step under the same execution semantics.
The validation agent replays and verifies this chain within the binary by confirming the existence of each propagation step and verifying exploitability under the defined threat model.
This process follows a validation-oriented task specification (Listing~\ref{lst:validation-prompt}), which constrains execution to the given evidence structure while verifying its correctness.

Each validation task thus operates as an evidence-driven replay, where execution is constrained by the structure of the input evidence chain.
Each validation task produces one of two outcomes—
(1) a verified, address-corrected propagation path if all evidence is consistent, or
(2) a rejection with explicit reasoning (e.g., the path does not exist, data is sanitized, or the sink is unreachable).
By grounding the validation process in previously recorded artifacts, FoA achieves replayable verification without requiring re-discovery or manual intervention.
After validation, the verified vulnerability and its associated evidence chains are aggregated into structured reports (List~\ref{lst:validated-evidence}).

\section{Evaluation}
\label{sec:exp}

In this section, we evaluate {\tech}'s effectiveness in real-world vulnerability detection tasks. 
Specifically, we aim to answer the following research questions:

\noindent\textbf{RQ1 (Effectiveness of Execution Model):}
Does FoA improve long-horizon vulnerability analysis compared to existing paradigms, including static analysis tools and prior LLM-based approaches?

\noindent\textbf{RQ2 (Mechanism Analysis):}
What are the key system mechanisms in FoA that contribute to its effectiveness in complex vulnerability analysis tasks?

\noindent\textbf{RQ3 (Efficiency and Scalability):}
Is FoA efficient and scalable for real-world vulnerability analysis in terms of cost, resource usage, and workload complexity?

\subsection{Experimental Settings}
\label{sec:sub:settings}

For the LLM backend, we use DeepSeek-v3 for vulnerability detection. 
To ensure deterministic outputs, we fix the temperature parameter at 0. 
Each agent is allowed up to 30 iterative reasoning steps per task instance, after which the process is terminated if no solution is found.

\textbf{Datasets}.
We evaluate {\tech} on 3,457 binaries extracted from real-world firmware images of four major IoT vendors (NETGEAR, D-Link, TP-Link, and Tenda), originally collected from the Karonte dataset~\cite{redini2020karonte}.
This dataset—also used by Mango~\cite{gibbs2024operation} and SaTC~\cite{chen2021sharing}—covers diverse device families and architectures commonly found in consumer IoT products.
Each binary satisfies both of the following conditions: (1) it contains at least one known dangerous function (sink); and (2) it contains at least one taint source from five predefined entry types: network, environment variable (including NVRAM and ENV), file, command-line argument (argv), and other. 
This ensures that all analyzed binaries are both security-relevant and comparable across different approaches.


\textbf{Baselines}. 
We compare {\tech} with four baselines.
(1) \textit{Operation Mango}~\cite{gibbs2024operation} is a SOTA tool for finding binary vulnerabilities via taint tracing and symbolic execution.
(2) SaTC~\cite{chen2021sharing} is a static analysis tool that leverages shared input keywords across firmware to detect bugs/vulnerabilities in embedded systems.
(3) LATTE~\cite{liu2025llm} is an LLM-powered static binary taint analysis system that integrates LLMs for taint analysis and vulnerability discovery in binaries and source code. 
(4) \textit{SWE Agent}~\cite{yang2024swe} is a framework that interleaves reasoning and tool actions in multiple steps, enabling dynamic decision-making and context tracking for software repositories. 
For fair comparison, we equip SWE Agent with the same set of tools and prompts as our system.

\textbf{Metrics}. 
We evaluate the performance of {\tech} using two key metrics that directly correspond to its two phases: \textit{discovery} and \textit{validation}.
\textit{(1) Underlying Vulnerability (Discovery Phase).}  
This refers to any finding automatically identified by the system during the discovery phase—i.e., a potential unsanitized data flow from a plausible untrusted source to a critical sink.
These findings represent candidate vulnerabilities inferred through LLM-based semantic reasoning and evidence collection.
\textit{(2) Verified Vulnerability (Validation Phase).} 
This refers to underlying vulnerabilities that were further confirmed as exploitable during the validation phase through concrete evidence replay.
When full manual labeling is unavailable, we estimate the total number of verified vulnerabilities by applying the measured precision from our random manual sample to the total underlying count.

\begin{table}[!t] \small
\centering
\caption{ Comparison of vulnerability detection between SaTC and {\tech}.}
\label{tab:satc}
\begin{tabular}{l|c c}
\toprule
 & SaTC   &  {\tech}\\
\midrule
CI            & 144 &  667 \\
BoF        & 44  & 329 \\
Affected Unique Binaries    & 131 & 591 \\
\bottomrule
\end{tabular}
\end{table}

\subsection{Effectiveness (RQ1)}
\label{sec:rq1}

We evaluate whether the FoA execution model improves both the effectiveness and practicality of vulnerability analysis.
We compare FoA against traditional static analysis tools and prior LLM-based approaches in terms of vulnerability detection effectiveness.

\textbf{Comparison with Static Analysis Paradigms.}
Table~\ref{tab:performance} summarizes the number of detected underlying vulnerabilities (“Vulns”), affected binaries, buffer overflows (“BoF”), and command injections (“CI”) across vendors for both {\tech} and \textit{Operation Mango}. Table~\ref{tab:satc} shows complementary results from SaTC, which focuses on different aspects of vulnerability analysis.
Although {\tech} reports fewer underlying vulnerabilities than \textit{Mango}, this difference reflects a fundamental distinction in execution behavior.
Traditional tools such as \textit{Mango} rely on large-scale path enumeration, which tends to produce many syntactic candidates, a substantial fraction of which cannot be validated.
In practice, the raw count of underlying alerts is less meaningful: a high number of alerts often corresponds to a large fraction of false positives, which in turn imposes a substantial verification burden on human analysts.

\begin{table}[!t]\small
\centering
\caption{Vulnerability detection prediction in {\tech} by type}
\label{tab:precision}
\begin{tabular}{l|ccc}
	\toprule
Vulnerability Type &  \makecell[c]{Underlying \\ Vulns} & \makecell[c]{Verified \\ Vulns} & Precision (\%) \\
\midrule
BoF    & 150 & 89 & 59.3 \\
CI     & 150 & 121  & 80.6 \\
Others & 50  & 34  & 68.0 \\
\bottomrule
\end{tabular}
\end{table}

\begin{table}[!t] \small
\centering
\caption{Comparison of estimated verified vulnerabilities for Operation Mango and {\tech}.}
\label{tab:valid-poc-comparison}
\begin{tabular}{l|cc|cc}
	\toprule
 & \multicolumn{2}{c|}{ Mango} & \multicolumn{2}{c}{{\tech}} \\
 & Total & Est. Valid (\%) & Total & Est. Valid (\%) \\
\midrule
BoF    & 1507 & 618 (41.0\%) & 329 & 195 (59.3\%) \\
CI     & 510  & 245 (48.1\%) & 667 & 538 (80.6\%) \\
Others & N/A  & N/A          & 278 & 189 (68.0\%) \\
\midrule
Total  & 2017 & $\sim$863 (42.7\%) & 1274 & $\approx$922 (72.3\%) \\
\bottomrule
\end{tabular}
\\
\footnotesize{\textit{Est. Valid (\%) indicates the estimated precision, i.e., the proportion of verified vulnerabilities among all underlying vulnerabilities.}}
\end{table}

This difference becomes more evident when considering validation outcomes.
The validation phase evaluates the exploitability of candidate vulnerabilities identified in discovery.
We randomly sampled 150 buffer overflow (BoF), 150 command injection (CI), and 50 other cases, and verified each vulnerability through independent review by two annotators.
A vulnerability was labeled as \texttt{verified} only if both confirmed an evidence-backed dataflow from an attacker-controlled input to a sensitive sink.
Table~\ref{tab:precision} shows that {\tech} achieves an overall precision of 72.3\%, significantly higher than the estimated 42.7\% for \textit{Mango}.
As shown in Table~\ref{tab:valid-poc-comparison}, this results in approximately 922 verified vulnerabilities for {\tech}, compared to $\sim$863 for \textit{Mango}.
Despite generating fewer raw alerts, {\tech} achieves comparable or higher numbers of verified vulnerabilities, demonstrating that FoA improves the quality of exploration rather than merely increasing its volume.
This behavior is consistent with the FoA design, where exploration and validation are tightly coupled.

\textbf{Comparison with LLM-based Paradigms.}
We further compare {\tech} with representative LLM-based approaches.
LATTE detects 219 vulnerabilities (94 unique), while SWE detects only 82 underlying vulnerabilities (11 verified).
LATTE follows a static pipeline where LLMs are used only for post-processing, limiting flexibility and coverage.
SWE adopts a single-agent, sequential reasoning model, which struggles to handle long and complex dataflow chains.

In contrast, FoA enables dynamic multi-agent generation and parallel exploration across multiple reasoning paths.
This allows {\tech} to scale to complex binaries and significantly improve both coverage and verified vulnerability yield.
The key difference lies not in the use of LLMs themselves, but in how reasoning is structured and executed.
These results suggest that simply incorporating LLMs is insufficient; instead, the execution model—particularly parallelization, decomposition, and evidence-driven validation—plays a critical role in effective analysis.

\begin{table}[!t] \small
\centering
\caption{Comparison of unique vulnerability types discovered by {\tech} and Mango.}
\label{tab:cwe-compare}
\begin{tabular}{l|c|c}
\toprule
     & \makecell[c]{Vulns \\ Type} & CWE-ID (Num.) \\
\midrule
\makecell[c]{Operation \\ Mango} & 2 & CWE-78 (510), CWE-120 (1507) \\
{\tech} & 6+ & \makecell[c]{CWE-22 (25),CWE-73 (61),\\ CWE-78 (667),CWE-120 (329),\\ CWE-134 (109), CWE-200 (73)} \\
\bottomrule
\end{tabular}
\end{table}

\textbf{Generalization and Robustness.}
FoA also improves generalization and robustness for binary vulnerability detection.
Table~\ref{tab:cwe-compare} shows that {\tech} supports a broader range of vulnerability types (6+ categories) compared to \textit{Mango}'s two primary types. Extending Mango to handle additional vulnerability categories requires manually designing new detection rules, integrating them into its static analysis pipeline, and performing additional engineering efforts. 
This broader coverage comes from agents inspecting dynamic intermediate artifacts (assembly, pseudocode, xrefs, strings) and using semantic understanding to judge diverse patterns.
This is consistent with FoA’s ability to adapt to diverse patterns without relying on manually crafted rules.
Such flexibility is enabled by its compositional execution structure, where agents dynamically interpret intermediate artifacts rather than relying on fixed rule templates.

\begin{table}[!t] \small
\centering
\caption{Vulnerability discovery by {\tech} in binaries where Operation Mango failed.}
\label{tab:failure-cases}
\begin{tabular}{l|r|r|r|r|r}
\toprule
Vendor & \makecell[r]{Angr\\ Errors} & \makecell[c]{OOM\\Kill} & \makecell[c]{Time\\out} & \makecell[c]{Affected \\ Binaries} & \makecell[r]{Underlying \\ Vulns} \\
\midrule
Netgear & 151 & 43 & 0 & 57 & 124 \\
Tenda & 13 & 5 & 3 & 10 & 25 \\
D-link & 67 & 17 & 0 & 14 & 23 \\
TP-link & 68 & 37 & 1 & 4 & 5 \\
\textbf{Total} & \textbf{299} & \textbf{102} & \textbf{4} & \textbf{85} & \textbf{179} \\
\bottomrule
\end{tabular}
\end{table}

{\tech} also identifies substantially more unique vulnerable binaries (591 vs.\ 244) and achieves a more balanced distribution across vendors.
Furthermore, Table~\ref{tab:failure-cases} shows that {\tech} successfully analyzes binaries where \textit{Mango} fails due to path explosion, memory exhaustion, or timeouts, discovering additional vulnerabilities in these challenging cases.
For example, {\tech} detects 429 affected binaries for Netgear (2.4$\times$ that of \textit{Mango}) and 61 for Tenda (nearly 3$\times$). 
In contrast, \textit{Mango}’s vulnerability findings are heavily skewed toward a small subset of binaries—mainly those from Netgear and Tenda—while its detection results for other vendors such as D-Link and TP-Link remain sparse.
This demonstrates that FoA improves not only detection capability but also robustness in real-world scenarios.

\textbf{Summary.}
Overall, the results consistently show that FoA improves long-horizon vulnerability analysis across multiple dimensions.
Rather than increasing the number of raw alerts, FoA restructures the execution process to produce higher-quality, verifiable vulnerabilities, while achieving broader coverage and better robustness than both traditional static analysis tools and prior LLM-based approaches.

\subsection{Mechanism Analysis (RQ2)}
\label{sec:rq2}

\begin{figure}[!t]
\centering
\includegraphics[width=2.4in]{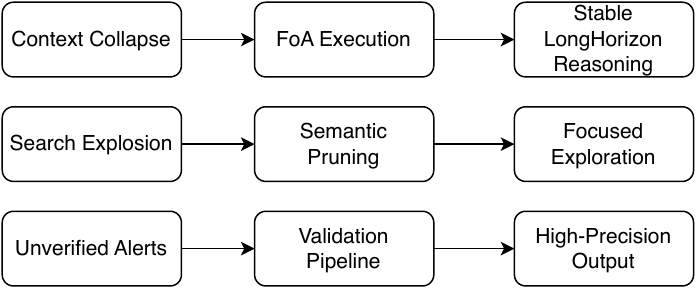}
\caption{FoA addresses three fundamental failure modes in long-horizon vulnerability analysis.}
\label{fig:foa-mechanism}
\end{figure}

To understand why FoA improves vulnerability analysis, we analyze the system through the lens of three fundamental failure modes in long-horizon reasoning: (1) \emph{context collapse} in linear execution, (2) \emph{search explosion} due to uncontrolled branching, and (3) \emph{unverified alerts} caused by the lack of systematic validation.
As illustrated in Figure~\ref{fig:foa-mechanism}, {\tech} addresses these challenges through three corresponding mechanisms: FoA execution model, LLM-guided semantic pruning, and a discovery–validation pipeline. 
We evaluate these mechanisms using controlled ablations on a sampled subset of 500 binaries from the Karonte dataset, under identical infrastructure and LLM settings.
Each experiment is repeated 5 times to ensure stable and comparable results.

\begin{table}[!t] \small
\centering
\caption{Ablation: Single Agent, Sequential-only Generation and {\tech}.}
\label{tab:swe-vs-ours}
\begin{tabular}{l|cc}
    \toprule
     & Vul. & Verified Vulns \\
\midrule
Single Agent & 8.4   & 1.3 \\
Sequential-only Generation & 45.8 & 22.3 \\
{\tech}      & 172.0 & 136.0 \\
\bottomrule
\end{tabular}
\end{table}

\textbf{Execution Structure: Mitigating Context Collapse.}
We first examine how FoA addresses \emph{context collapse}, a common failure mode in long-horizon reasoning where the LLM fails to maintain extended dependency chains.
We compare {\tech} with two degraded variants: a single-agent system and a sequential-only execution model.
As shown in Table~\ref{tab:swe-vs-ours}, the single-agent variant detects only 8.4 underlying vulnerabilities and 1.3 verified vulnerabilities, while sequential-only execution improves to 45.8 and 22.3, respectively, but remains far below {\tech} (172.0 / 136.0).
These results indicate that linear reasoning structures cannot sustain deep and branching dataflow analysis.
In the single-agent setting, the LLM must maintain the entire reasoning state within a single context, leading to rapid degradation as analysis depth increases.
Sequential execution partially alleviates this issue but still enforces strict serialization, causing early context loss and limiting coverage.

In contrast, FoA decomposes reasoning into dynamically generated agents and enables parallel exploration across multiple paths.
This design localizes reasoning contexts and preserves intermediate evidence, effectively mitigating context collapse and enabling scalable long-horizon analysis.

\textbf{Search Control: Mitigating Search Explosion.}
We next analyze how FoA addresses \emph{search explosion}, where uncontrolled branching leads to excessive and unproductive exploration.
Table~\ref{tab:pruning-ablation} compares full {\tech} with a NoPrune variant that enforces exhaustive exploration.
While NoPrune increases the number of agents (40.4 vs.\ 34.3) and reasoning steps (503.1 vs.\ 464.1), it reduces verified vulnerabilities from 136.0 to 98.7.
This demonstrates that naive expansion of the search space is counterproductive: without filtering, the system explores many semantically irrelevant paths, diluting computational resources and increasing noise.

FoA addresses this issue through LLM-guided semantic pruning, which selectively expands only meaningful branches.
Although LLMs exhibit some implicit filtering ability, such filtering is coarse-grained and inconsistent.
The explicit pruning mechanism provides fine-grained control over exploration, ensuring efficient and focused reasoning.
The degradation in verified vulnerabilities under NoPrune directly attributes performance gains to the pruning mechanism, confirming its causal role in controlling search explosion.

\begin{table}[!t]\small
\centering
\caption{Pruning ablation between full {\tech} and NoPrune. }
\label{tab:pruning-ablation}
\begin{tabular}{p{1.6cm}|p{0.7cm}|p{1cm}|p{1.5cm}|p{1.3cm}}
\toprule
 & Vul. & Verified  & Avg agents & Avg steps  \\
\midrule
{\tech}          & 172.2 & 136.0  & 34.3 & 464.1 \\
NoPrune & 167.4 & 98.7  & 40.4 & 503.1  \\
\midrule
Failure rate & \multicolumn{4}{c}{Full: 0.00\% \quad NoPrune: 0.00\%} \\
\bottomrule
\end{tabular}
\end{table}

\textbf{Validation Pipeline: Mitigating Unverified Alerts.}
We finally analyze how FoA addresses \emph{unverified alerts}, a common issue where systems produce large numbers of candidate vulnerabilities without confirming exploitability.
FoA adopts a two-stage pipeline consisting of discovery and validation.
The discovery stage generates candidate vulnerabilities, while the validation stage verifies exploitability through evidence-backed analysis.
The effect of this design is reflected in the gap between underlying and verified vulnerabilities reported in RQ1, which quantifies the number of spurious candidates filtered out during validation.
Without this stage, the system would resemble traditional static analysis tools that produce many unverifiable alerts.
The validation mechanism thus ensures that reported vulnerabilities are actionable, significantly improving precision.
This separation between discovery and validation stages provides an explicit mechanism for eliminating false positives, linking the observed precision improvements in RQ1 to the validation pipeline design.

\textbf{Summary.}
Overall, FoA's effectiveness can be understood as addressing three failure modes in long-horizon vulnerability analysis: context collapse, search explosion, and unverified alerts.
Through controlled ablations, we show that each mechanism contributes causally to performance improvements by isolating and mitigating these failure modes.
In short, FoA transforms vulnerability analysis from linear reasoning into a structured and robust process.

\subsection{Efficiency and Scalability (RQ3)}
\label{sec:rq3}

We evaluate whether {\tech} is efficient and scalable for real-world vulnerability analysis from three perspectives: 
(1) end-to-end analysis cost per binary, 
(2) cost normalized by useful outputs (verified vulnerabilities), and 
(3) scalability under varying workload complexity.

\textbf{End-to-End Efficiency.}  
Table~\ref{tab:scalability} summarizes the average per-binary cost of {\tech}. 
On average, {\tech} completes analysis within 43.8 minutes per binary, consuming 1.61M tokens (1.29M for discovery and 0.32M for validation).  
Each run involves 33.8 agents and 464 reasoning steps.
These results indicate that {\tech} maintains moderate per-instance cost despite extensive exploration. 
Notably, the large number of agents does not lead to prohibitive overhead, suggesting that the lightweight agent abstraction enables parallel exploration without significantly increasing marginal cost.

\textbf{Cost per Verified Vulnerability.}  
Since different systems produce substantially different numbers of findings, raw execution cost is not directly comparable. 
We therefore normalize cost by the number of verified vulnerabilities.
As shown in Table~\ref{tab:per_verified_cost}, {\tech} requires 140.2 minutes and 4.71M tokens per verified vulnerability, compared to 357.1 minutes and 12.5M tokens for the single-agent baseline.  
This corresponds to an approximate 2.5$\times$ improvement in both time and token efficiency.
This gain stems from {\tech}'s two-stage execution model: the discovery phase aggressively explores candidate paths, while the validation phase filters and confirms only high-confidence findings. 
As a result, computational resources are concentrated on actionable outputs rather than redundant reasoning trajectories.

\begin{table}[!t] \small
\centering
\caption{The overview of time cost and token usage in {\tech}. per binary}
\label{tab:scalability}
\begin{tabular}{l|c}
\toprule
             & Value \\
\midrule
Average time  & 43.8 min \\
Average tokens (discovery / validation) &  1.29 / 0.32 (M) \\
Average agents  & 33.79 \\
Average reasoning steps & 464.25 \\
\bottomrule
\end{tabular}
\end{table}

\begin{table}[!t]\small
\centering
\caption{Per-verified vulnerability cost and time.}
\label{tab:per_verified_cost}
\begin{tabular}{l|c c }
\toprule
Variant & Time/vul.  & Tokens/vul.  \\
\midrule
Single Agent &  357.1 min &  12.5 M  \\
{\tech}  &  140.2 min &  4.71 M\\
\bottomrule
\end{tabular}
\end{table}

\begin{figure}
    \centering
    \includegraphics[width=0.8\linewidth]{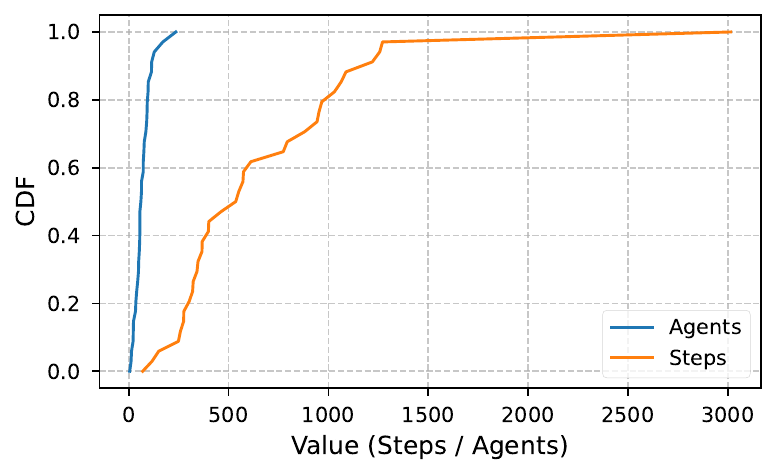}
  \caption{CDF of reasoning steps and the number of agents created per in the discovery phase.}
    \label{fig:placeholder}
\end{figure}

\textbf{Scalability with Analysis Complexity.}  
To understand how {\tech} scales with workload complexity, we analyze the distribution of reasoning steps and agent usage in Figure~\ref{fig:placeholder}.  
Both exhibit a long-tail distribution: most binaries require moderate effort, while a small fraction demands significantly deeper exploration.
Importantly, resource consumption (tokens and time cost) correlates primarily with reasoning depth and path complexity, rather than the number of instantiated agents. 
This behavior suggests that {\tech} scales with intrinsic task complexity rather than suffering from systemic inefficiencies, making it suitable for heterogeneous real-world workloads.

\textbf{Scalability in Large-Scale Deployment.}  
We further evaluate {\tech} on 3,457 real-world binaries across multiple vendors.  
{\tech} successfully analyzes all binaries and identifies a substantially larger number of vulnerabilities compared to prior systems, while maintaining manageable per-instance cost.
This demonstrates that {\tech} can sustain both high throughput and high yield in large-scale settings, a key requirement for practical deployment.

\textbf{Summary.}
{\tech} achieves efficiency–effectiveness trade-offs, reducing cost per verified vulnerability while scaling to complex and large-scale analysis workloads.
\section{Discussion and Limitations}
\label{sec:discuss}

\textbf{Stability and Variance in LLM-driven Execution.}
A potential concern is the stability of {\tech} across runs and inputs.
While we reduce randomness by setting temperature to zero and enforcing evidence-based reasoning constraints, the system may still exhibit variability due to the inherent non-determinism of LLMs.
Importantly, this variability manifests primarily at the reasoning path level (e.g., different exploration trajectories), rather than completely invalid outputs.
Our empirical results show that, despite such variations, the overall effectiveness remains stable across repeated runs and diverse binary sets.
This suggests that FoA mitigates—but does not fully eliminate—LLM-induced variance by structuring reasoning into constrained and verifiable steps.

\textbf{Residual Risk of Hallucination.}
Although FoA enforces evidence-backed reasoning and validation, hallucination cannot be fully eliminated.
In particular, errors may still arise when intermediate representations (e.g., decompiled code or dataflow traces) are ambiguous or incomplete.
To mitigate this, {\tech} requires structured outputs, enforces schema validation, and applies independent verification for sampled cases (Table~\ref{tab:precision}).
However, these safeguards rely on the availability and correctness of supporting evidence.
As a result, hallucination remains a residual risk, especially for complex binaries with unclear semantics.

\textbf{Dependence on Underlying Analysis Tools.}
{\tech} relies on external analysis tools (currently Radare2Tool) to provide intermediate artifacts such as disassembly, control flow, and cross-references.
Limitations of these tools—such as inaccurate function boundary recovery, unresolved indirect calls, or aliasing issues—directly affect the quality of downstream reasoning.
While FoA partially mitigates these issues by reasoning over multiple intermediate representations and refining hypotheses iteratively, it cannot recover information that is fundamentally missing or incorrect.
Therefore, the effectiveness of {\tech} is bounded by the fidelity of the underlying analysis tools.
Future work may explore integrating complementary techniques such as symbolic execution or fuzzing to improve analysis completeness.

\textbf{Incomplete Coverage.}
Despite improved exploration efficiency, {\tech} does not guarantee full path coverage.
Binary analysis inherently suffers from path explosion, complex control flows, and indirect jumps.
FoA alleviates this issue through structured exploration and pruning (RQ2), but may still miss vulnerabilities located in rarely explored or highly obfuscated paths.
This limitation reflects a fundamental trade-off between exploration completeness and computational tractability.
Improving coverage likely requires hybrid approaches that combine static, dynamic, and learning-based techniques.

\textbf{Scope of Mechanism Validation.}
Finally, while our ablation study (RQ2) isolates key mechanisms—execution structure, pruning, and validation—it is conducted on a sampled subset of binaries.
Although the results are consistent and repeated across runs, they may not capture all edge cases in large-scale real-world deployments.
Extending mechanism-level validation to broader datasets and more diverse environments remains an important direction for future work.
\section{Related Work}
\label{sec:related}

\sloppy

\textbf{Binary Analysis}.
Static binary analysis includes rule-based detection (e.g., Flawfinder~\cite{flawfinder}, CWE-checker~\cite{cwe-checker}) and symbolic execution (e.g., Mango~\cite{gibbs2024operation}, SaTC~\cite{chen2021sharing}, Karonte~\cite{redini2020karonte}). 
These approaches improve coverage and precision through techniques such as data-flow analysis, constraint solving, and path exploration. 
Dynamic approaches, including fuzzing~\cite{feng2021snipuzz, li2022muafl, scharnowski2022fuzzware, zheng2019firm} and concolic execution~\cite{coppa2022symfusion, poeplau2020symbolic, stephens2016driller, yun2018qsym}, complement static analysis by exploring runtime behaviors, but rely heavily on execution environments and input generation strategies. 
In practice, both paradigms require substantial engineering effort to balance coverage and efficiency, especially in complex firmware settings. 
Despite these advances, existing binary analysis techniques largely operate under a one-pass or weakly adaptive paradigm, where exploration strategies are fixed or only locally adjusted, limiting the ability to incorporate intermediate results into global decision making and leading to path explosion, limited semantic adaptability, and high engineering cost.

\textbf{Systems for Multi-Path and Tool-Mediated Execution}.
Several works study how to scale multi-path exploration and tool--computation loops under resource constraints~\cite{chipounov2011s2e, yao2022react, schick2023toolformer, mei2025helix, stoica2024retry, wu2024oscopilotgeneralistcomputeragents, xie2024osworldbenchmarkingmultimodalagents, yang2023intercodestandardizingbenchmarkinginteractive}. 
These systems typically treat complex tasks as iterative processes that interleave reasoning, execution, and feedback, enabling adaptive decision making during execution. 
For example, S2E~\cite{chipounov2011s2e} demonstrates selective symbolic execution with execution multiplexing, while more recent LLM-based systems emphasize closed-loop interaction through tool use and environment feedback. 
At larger scales, LLM systems are integrated into infrastructure and pipelines, such as scheduling in heterogeneous environments~\cite{mei2025helix} and debugging workflows~\cite{stoica2024retry}, highlighting the role of resource-aware execution. 
However, these systems primarily focus on scalability, interaction patterns, or resource management, and do not provide a structured execution model for coordinating long-horizon reasoning, exploration, and validation in semantically complex domains such as binary analysis.

\textbf{LLM Agents and LLM-assisted Code Analysis}.
LLMs demonstrate strong reasoning capabilities, enhanced by prompting strategies such as task decomposition, chain-of-thought, tree-of-thought, and self-reflection~\cite{wang2022self, wei2022chain, yao2023tree, shinn2023reflexion, liu2023languages}. 
LLM agents extend these capabilities into tool-augmented systems~\cite{park2023generative, wang2024survey, xi2023rise}, enabling iterative reasoning with external tools and environments. 
These approaches have been applied to software engineering~\cite{copilot} and security tasks~\cite{chen2023teaching, feng2023prompting, pei2023can, pearce2022pop}, including bug reproduction~\cite{feng2023prompting}, vulnerability discovery~\cite{li2023hitchhiker, pearce2023examining}, and test generation~\cite{deng2023large, lemieux2023codamosa}. 
Recent systems such as LLMSAN~\cite{wang2024sanitizing} and LATTE~\cite{liu2025llm} further explore applying LLMs to program and binary analysis. 
However, existing LLM-based approaches primarily improve reasoning quality at the prompt or pipeline level, while their execution remains largely linear or weakly structured, making them prone to context collapse, uncontrolled exploration, and lack of systematic validation in long-horizon tasks.

In contrast, FoA introduces a structured execution model that unifies reasoning, exploration, and validation, systematically addressing context collapse, search explosion, and unverified alerts in long-horizon vulnerability analysis.

\fussy 
\section{Conclusion}
\label{sec:conclusion}

\sloppy

We proposed {\tech}, an LLM-driven system that rethinks binary vulnerability analysis as an iterative and structured execution process rather than a one-pass pipeline. 
By tightly interleaving reasoning and analysis through a reasoning–action–observation loop, {\tech} enables adaptive exploration and incremental construction of semantic evidence.
To support scalable long-horizon analysis, we introduce a FoA execution model, which decomposes tasks and coordinates parallel exploration while maintaining bounded reasoning contexts. 
We evaluate {\tech} on 3,457 real-world binaries and show that it identifies 1,274 vulnerabilities across 591 unique binaries with 72.3\% precision, covering a broader range of vulnerability types than prior approaches.
Overall, our results demonstrate that restructuring vulnerability analysis as a coordinated, multi-agent execution process—rather than improving individual analysis components—provides a principled way to address context collapse, search explosion, and unverified alerts, enabling more effective and scalable binary vulnerability detection.

\fussy 

 \clearpage
\bibliographystyle{ACM-Reference-Format}
\bibliography{ref/ref_code, ref/ref_online, ref/ref_security, ref/ref_firmware, ref/ref_llm, ref/ref_systems, ref/ref_ours}

@article{yang2024swe,
  title={Swe-agent: Agent-computer interfaces enable automated software engineering},
  author={Yang, John and Jimenez, Carlos E and Wettig, Alexander and Lieret, Kilian and Yao, Shunyu and Narasimhan, Karthik and Press, Ofir},
  journal={Advances in Neural Information Processing Systems},
  volume={37},
  pages={50528--50652},
  year={2024}
}

@article{jimenez2023swe,
  title={Swe-bench: Can language models resolve real-world github issues?},
  author={Jimenez, Carlos E and Yang, John and Wettig, Alexander and Yao, Shunyu and Pei, Kexin and Press, Ofir and Narasimhan, Karthik},
  journal={arXiv preprint arXiv:2310.06770},
  year={2023}
}

@inproceedings{chen2016towards,
  title={Towards automated dynamic analysis for linux-based embedded firmware.},
  author={Chen, Daming D and Woo, Maverick and Brumley, David and Egele, Manuel},
  booktitle={NDSS},
  year={2016}
}

@article{wu2023autogen,
  title={AutoGen: Enabling next-gen LLM applications via multi-agent conversation framework},
  author={Wu, Yizhou and Yang, Diyi and Wang, Kexin and Tan, Vincent Y. F. and Xu, Canwen and Yang, Zhenyu and Li, Xiang and Tan, Xiaoxiao and others},
  journal={arXiv preprint arXiv:2309.12307},
  year={2023}
}

@misc{hussain2025vulbinllm,
  title={VulBinLLM: LLM-powered Vulnerability Detection for Stripped Binaries},
  author={Hussain, Nasir and Chen, Haohan and Tran, Chanh and Huang, Philip and Li, Zhuohao and Chugh, Pravir and Chen, William and Kundu, Ashish and Tian, Yuan},
  year={2025},
  eprint={2505.22010},
  archivePrefix={arXiv},
  primaryClass={cs.SE},
  url={https://arxiv.org/abs/2505.22010}
}

@inproceedings{chen2025clearagent,
  title={ClearAgent: Agentic Binary Analysis for Effective Vulnerability Detection},
  author={Chen, Xiang and Ye, Chengfeng and Zhou, Anshunkang and Zhang, Charles},
  booktitle={Proceedings of the 1st ACM SIGPLAN International Workshop on Language Models and Programming Languages (LMPL 2025), co-located with ICFP/SPLASH 2025},
  pages={130--137},
  year={2025},
  address={Singapore},
  publisher={ACM},
  doi={10.1145/3759425.3763397}
}

@article{liu2025llm,
  title={Llm-powered static binary taint analysis},
  author={Liu, Puzhuo and Sun, Chengnian and Zheng, Yaowen and Feng, Xuan and Qin, Chuan and Wang, Yuncheng and Xu, Zhenyang and Li, Zhi and Di, Peng and Jiang, Yu and others},
  journal={ACM Transactions on Software Engineering and Methodology},
  volume={34},
  number={3},
  pages={1--36},
  year={2025},
  publisher={ACM New York, NY}
}

@inproceedings{pearce2023examining,
  title={Examining zero-shot vulnerability repair with large language models},
  author={Pearce, Hammond and Tan, Benjamin and Ahmad, Baleegh and Karri, Ramesh and Dolan-Gavitt, Brendan},
  booktitle={2023 IEEE Symposium on Security and Privacy (SP)},
  pages={2339--2356},
  year={2023},
  organization={IEEE}
}

@article{chen2023teaching,
  title={Teaching large language models to self-debug},
  author={Chen, Xinyun and Lin, Maxwell and Sch{\"a}rli, Nathanael and Zhou, Denny},
  journal={arXiv preprint arXiv:2304.05128},
  year={2023}
}

@article{feng2023prompting,
  title={Prompting Is All Your Need: Automated Android Bug Replay with Large Language Models},
  author={Feng, Sidong and Chen, Chunyang},
  journal={arXiv preprint arXiv:2306.01987},
  year={2023}
}

@inproceedings{pei2023can,
  title={Can large language models reason about program invariants?},
  author={Pei, Kexin and Bieber, David and Shi, Kensen and Sutton, Charles and Yin, Pengcheng},
  booktitle={International Conference on Machine Learning},
  pages={27496--27520},
  year={2023},
  organization={PMLR}
}

@article{yao2023tree,
  title={Tree of thoughts: Deliberate problem solving with large language models},
  author={Yao, Shunyu and Yu, Dian and Zhao, Jeffrey and Shafran, Izhak and Griffiths, Thomas L and Cao, Yuan and Narasimhan, Karthik},
  journal={arXiv preprint arXiv:2305.10601},
  year={2023}
}

@article{wei2022chain,
  title={Chain-of-thought prompting elicits reasoning in large language models},
  author={Wei, Jason and Wang, Xuezhi and Schuurmans, Dale and Bosma, Maarten and Xia, Fei and Chi, Ed and Le, Quoc V and Zhou, Denny and others},
  journal={Advances in Neural Information Processing Systems},
  volume={35},
  pages={24824--24837},
  year={2022}
}

@article{wang2022self,
  title={Self-consistency improves chain of thought reasoning in language models},
  author={Wang, Xuezhi and Wei, Jason and Schuurmans, Dale and Le, Quoc and Chi, Ed and Narang, Sharan and Chowdhery, Aakanksha and Zhou, Denny},
  journal={arXiv preprint arXiv:2203.11171},
  year={2022}
}

@inproceedings{shinn2023reflexion,
  title={Reflexion: Language agents with verbal reinforcement learning},
  author={Shinn, Noah and Cassano, Federico and Gopinath, Ashwin and Narasimhan, Karthik R and Yao, Shunyu},
  booktitle={Thirty-seventh Conference on Neural Information Processing Systems},
  year={2023}
}

@misc{wu2024oscopilotgeneralistcomputeragents,
      title={OS-Copilot: Towards Generalist Computer Agents with Self-Improvement}, 
      author={Zhiyong Wu and Chengcheng Han and Zichen Ding and Zhenmin Weng and Zhoumianze Liu and Shunyu Yao and Tao Yu and Lingpeng Kong},
      year={2024},
      eprint={2402.07456},
      archivePrefix={arXiv},
      primaryClass={cs.AI},
      url={https://arxiv.org/abs/2402.07456}, 
}

@misc{xie2024osworldbenchmarkingmultimodalagents,
      title={OSWorld: Benchmarking Multimodal Agents for Open-Ended Tasks in Real Computer Environments}, 
      author={Tianbao Xie and Danyang Zhang and Jixuan Chen and Xiaochuan Li and Siheng Zhao and Ruisheng Cao and Toh Jing Hua and Zhoujun Cheng and Dongchan Shin and Fangyu Lei and Yitao Liu and Yiheng Xu and Shuyan Zhou and Silvio Savarese and Caiming Xiong and Victor Zhong and Tao Yu},
      year={2024},
      eprint={2404.07972},
      archivePrefix={arXiv},
      primaryClass={cs.AI},
      url={https://arxiv.org/abs/2404.07972}, 
}

@misc{yang2023intercodestandardizingbenchmarkinginteractive,
      title={InterCode: Standardizing and Benchmarking Interactive Coding with Execution Feedback}, 
      author={John Yang and Akshara Prabhakar and Karthik Narasimhan and Shunyu Yao},
      year={2023},
      eprint={2306.14898},
      archivePrefix={arXiv},
      primaryClass={cs.CL},
      url={https://arxiv.org/abs/2306.14898}, 
}

@article{pearce2022pop,
  title={Pop Quiz! Can a Large Language Model Help With Reverse Engineering?},
  author={Pearce, Hammond and Tan, Benjamin and Krishnamurthy, Prashanth and Khorrami, Farshad and Karri, Ramesh and Dolan-Gavitt, Brendan},
  journal={arXiv preprint arXiv:2202.01142},
  year={2022}
}

@article{li2023hitchhiker,
  title={The Hitchhiker's Guide to Program Analysis: A Journey with Large Language Models},
  author={Li, Haonan and Hao, Yu and Zhai, Yizhuo and Qian, Zhiyun},
  journal={arXiv preprint arXiv:2308.00245},
  year={2023}
}

@article{yao2022react,
  title={React: Synergizing reasoning and acting in language models},
  author={Yao, Shunyu and Zhao, Jeffrey and Yu, Dian and Du, Nan and Shafran, Izhak and Narasimhan, Karthik and Cao, Yuan},
  journal={arXiv preprint arXiv:2210.03629},
  year={2022}
}

@article{schick2023toolformer,
  title={Toolformer: Language models can teach themselves to use tools},
  author={Schick, Timo and Dwivedi-Yu, Jane and Dess{\`\i}, Roberto and Raileanu, Roberta and Lomeli, Maria and Hambro, Eric and Zettlemoyer, Luke and Cancedda, Nicola and Scialom, Thomas},
  journal={Advances in Neural Information Processing Systems},
  volume={36},
  pages={68539--68551},
  year={2023}
}

@article{liu2023languages,
  title={Languages are rewards: Hindsight finetuning using human feedback},
  author={Liu, Hao and Sferrazza, Carmelo and Abbeel, Pieter},
  journal={arXiv preprint arXiv:2302.02676},
  year={2023}
}

@inproceedings{park2023generative,
  title={Generative agents: Interactive simulacra of human behavior},
  author={Park, Joon Sung and O'Brien, Joseph and Cai, Carrie Jun and Morris, Meredith Ringel and Liang, Percy and Bernstein, Michael S},
  booktitle={Proceedings of the 36th annual acm symposium on user interface software and technology},
  pages={1--22},
  year={2023}
}

@article{wang2024survey,
  title={A survey on large language model based autonomous agents},
  author={Wang, Lei and Ma, Chen and Feng, Xueyang and Zhang, Zeyu and Yang, Hao and Zhang, Jingsen and Chen, Zhiyuan and Tang, Jiakai and Chen, Xu and Lin, Yankai and others},
  journal={Frontiers of Computer Science},
  volume={18},
  number={6},
  pages={186345},
  year={2024},
  publisher={Springer}
}

@article{xi2023rise,
  title={The rise and potential of large language model based agents: A survey},
  author={Xi, Zhiheng and Chen, Wenxiang and Guo, Xin and He, Wei and Ding, Yiwen and Hong, Boyang and Zhang, Ming and Wang, Junzhe and Jin, Senjie and Zhou, Enyu and others},
  journal={arXiv preprint arXiv:2309.07864},
  year={2023}
}

@inproceedings{deng2023large,
  title={Large language models are zero-shot fuzzers: Fuzzing deep-learning libraries via large language models},
  author={Deng, Yinlin and Xia, Chunqiu Steven and Peng, Haoran and Yang, Chenyuan and Zhang, Lingming},
  booktitle={Proceedings of the 32nd ACM SIGSOFT international symposium on software testing and analysis},
  pages={423--435},
  year={2023}
}

@inproceedings{lemieux2023codamosa,
  title={Codamosa: Escaping coverage plateaus in test generation with pre-trained large language models},
  author={Lemieux, Caroline and Inala, Jeevana Priya and Lahiri, Shuvendu K and Sen, Siddhartha},
  booktitle={2023 IEEE/ACM 45th International Conference on Software Engineering (ICSE)},
  pages={919--931},
  year={2023},
  organization={IEEE}
}

@inproceedings{wang2024sanitizing,
  title={Sanitizing large language models in bug detection with data-flow},
  author={Wang, Chengpeng and Zhang, Wuqi and Su, Zian and Xu, Xiangzhe and Zhang, Xiangyu},
  booktitle={Findings of the Association for Computational Linguistics: EMNLP 2024},
  pages={3790--3805},
  year={2024}
}

@ONLINE{copilot, 
title = {{Copilot: The AI developer tool.}},
author={Microsoft Org.},
year ={2023},
Howpublished = {\url{https://github.com/features/copilot}}
}

@ONLINE{cwe-checker, title= {Detect common bug classes formally known as Common Weakness Enumerations (CWEs).},
author={Fraunhofer FKIE},
year={accessible 2024},
Howpublished ={\url{https://github.com/fkie-cad/cwe_checker}}
}

@ONLINE{flawfinder, title= {Flawfinder is a simple program that scans C/C++ source code and reports potential security flaws.},
author={David A. Wheeler},
year={accessible 2024},
Howpublished = {\url{https://github.com/david-a-wheeler/flawfinder}}
}

@inproceedings{chen2021sharing,
  title={Sharing more and checking less: Leveraging common input keywords to detect bugs in embedded systems},
  author={Chen, Libo and Wang, Yanhao and Cai, Quanpu and Zhan, Yunfan and Hu, Hong and Linghu, Jiaqi and Hou, Qinsheng and Zhang, Chao and Duan, Haixin and Xue, Zhi},
  booktitle={30th USENIX Security Symposium (USENIX Security 21)},
  pages={303--319},
  year={2021}
}

@inproceedings{redini2020karonte,
  title={Karonte: Detecting insecure multi-binary interactions in embedded firmware},
  author={Redini, Nilo and Machiry, Aravind and Wang, Ruoyu and Spensky, Chad and Continella, Andrea and Shoshitaishvili, Yan and Kruegel, Christopher and Vigna, Giovanni},
  booktitle={2020 IEEE Symposium on Security and Privacy (SP)},
  pages={1544--1561},
  year={2020},
  organization={IEEE}
}

@inproceedings{gibbs2024operation,
  title={Operation mango: Scalable discovery of $\{$Taint-Style$\}$ vulnerabilities in binary firmware services},
  author={Gibbs, Wil and Raj, Arvind S and Vadayath, Jayakrishna Menon and Tay, Hui Jun and Miller, Justin and Ajayan, Akshay and Basque, Zion Leonahenahe and Dutcher, Audrey and Dong, Fangzhou and Maso, Xavier and others},
  booktitle={33rd USENIX Security Symposium (USENIX Security 24)},
  pages={7123--7139},
  year={2024}
}

@inproceedings{christensen2020decaf,
  title={$\{$DECAF$\}$: Automatic, adaptive de-bloating and hardening of $\{$COTS$\}$ firmware},
  author={Christensen, Jake and Anghel, Ionut Mugurel and Taglang, Rob and Chiroiu, Mihai and Sion, Radu},
  booktitle={29th USENIX Security Symposium (USENIX Security 20)},
  pages={1713--1730},
  year={2020}
}

@inproceedings{busch2023teezz,
  title={Teezz: Fuzzing trusted applications on cots android devices},
  author={Busch, Marcel and Machiry, Aravind and Spensky, Chad and Vigna, Giovanni and Kruegel, Christopher and Payer, Mathias},
  booktitle={2023 IEEE Symposium on Security and Privacy (SP)},
  pages={1204--1219},
  year={2023},
  organization={IEEE}
}

@inproceedings{han2023queryx,
  title={Queryx: Symbolic query on decompiled code for finding bugs in COTS binaries},
  author={Han, HyungSeok and Kyea, JeongOh and Jin, Yonghwi and Kang, Jinoh and Pak, Brian and Yun, Insu},
  booktitle={2023 IEEE Symposium on Security and Privacy (SP)},
  pages={3279--3295},
  year={2023},
  organization={IEEE}
}

@inproceedings{feng2021snipuzz,
  title={Snipuzz: Black-box fuzzing of iot firmware via message snippet inference},
  author={Feng, Xiaotao and Sun, Ruoxi and Zhu, Xiaogang and Xue, Minhui and Wen, Sheng and Liu, Dongxi and Nepal, Surya and Xiang, Yang},
  booktitle={Proceedings of the 2021 ACM SIGSAC conference on computer and communications security},
  pages={337--350},
  year={2021}
}

@inproceedings{li2022muafl,
  title={$\mu$AFL: non-intrusive feedback-driven fuzzing for microcontroller firmware},
  author={Li, Wenqiang and Shi, Jiameng and Li, Fengjun and Lin, Jingqiang and Wang, Wei and Guan, Le},
  booktitle={Proceedings of the 44th International Conference on Software Engineering},
  pages={1--12},
  year={2022}
}

@inproceedings{scharnowski2022fuzzware,
  title={Fuzzware: Using precise $\{$MMIO$\}$ modeling for effective firmware fuzzing},
  author={Scharnowski, Tobias and Bars, Nils and Schloegel, Moritz and Gustafson, Eric and Muench, Marius and Vigna, Giovanni and Kruegel, Christopher and Holz, Thorsten and Abbasi, Ali},
  booktitle={31st USENIX Security Symposium (USENIX Security 22)},
  pages={1239--1256},
  year={2022}
}

@inproceedings{zheng2019firm,
  title={$\{$FIRM-AFL$\}$:$\{$High-Throughput$\}$ greybox fuzzing of $\{$IoT$\}$ firmware via augmented process emulation},
  author={Zheng, Yaowen and Davanian, Ali and Yin, Heng and Song, Chengyu and Zhu, Hongsong and Sun, Limin},
  booktitle={28th USENIX Security Symposium (USENIX Security 19)},
  pages={1099--1114},
  year={2019}
}

@inproceedings{coppa2022symfusion,
  title={Symfusion: hybrid instrumentation for concolic execution},
  author={Coppa, Emilio and Yin, Heng and Demetrescu, Camil},
  booktitle={Proceedings of the 37th IEEE/ACM International Conference on Automated Software Engineering},
  pages={1--12},
  year={2022}
}

@inproceedings{poeplau2020symbolic,
  title={Symbolic execution with $\{$SymCC$\}$: Don't interpret, compile!},
  author={Poeplau, Sebastian and Francillon, Aur{\'e}lien},
  booktitle={29th USENIX Security Symposium (USENIX Security 20)},
  pages={181--198},
  year={2020}
}

@inproceedings{stephens2016driller,
  title={Driller: Augmenting fuzzing through selective symbolic execution.},
  author={Stephens, Nick and Grosen, John and Salls, Christopher and Dutcher, Andrew and Wang, Ruoyu and Corbetta, Jacopo and Shoshitaishvili, Yan and Kruegel, Christopher and Vigna, Giovanni},
  booktitle={NDSS},
  volume={16},
  number={2016},
  pages={1--16},
  year={2016}
}

@inproceedings{yun2018qsym,
  title={$\{$QSYM$\}$: A practical concolic execution engine tailored for hybrid fuzzing},
  author={Yun, Insu and Lee, Sangho and Xu, Meng and Jang, Yeongjin and Kim, Taesoo},
  booktitle={27th USENIX Security Symposium (USENIX Security 18)},
  pages={745--761},
  year={2018}
}

@inproceedings{chipounov2011s2e,
  author = {Chipounov, Vitaly and Kuznetsov, Volodymyr and Candea, George},
  title = {{S2E}: A Platform for In-Vivo Multi-Path Analysis of Software Systems},
  booktitle = {Proceedings of the 16th International Conference on Architectural Support for Programming Languages and Operating Systems},
  series = {ASPLOS XVI},
  year = {2011},
  pages = {265--278},
  address = {New York, NY, USA},
  publisher = {ACM},
  doi = {10.1145/1950365.1950396}
}

@inproceedings{mei2025helix,
  author = {Mei, Yixuan and Zhuang, Yonghao and Miao, Xupeng and Yang, Juncheng and Jia, Zhihao and Vinayak, Rashmi},
  title = {Helix: Serving Large Language Models over Heterogeneous {GPUs} and Network via Max-Flow},
  booktitle = {Proceedings of the 30th ACM International Conference on Architectural Support for Programming Languages and Operating Systems},
  series = {ASPLOS '25},
  year = {2025},
  address = {New York, NY, USA},
  publisher = {ACM},
  doi = {10.1145/3669940.3707215}
}

@inproceedings{stoica2024retry,
  author = {Stoica, Bogdan Alexandru and Sethi, Utsav and Su, Yiming and Zhou, Cyrus and Lu, Shan and Mace, Jonathan and Musuvathi, Madanlal and Nath, Suman},
  title = {If At First You Don't Succeed, Try, Try, Again\ldots? Insights and {LLM}-Informed Tooling for Detecting Retry Bugs in Software Systems},
  booktitle = {Proceedings of the 30th ACM Symposium on Operating Systems Principles},
  series = {SOSP '24},
  year = {2024},
  address = {New York, NY, USA},
  publisher = {ACM},
  doi = {10.1145/3694715.3695971}
}

\appendix

\section{System Implementation}

\begin{table}[htb]
\small
\centering
\caption{Tools Available to Agents.}
\label{tab:tools}
\begin{tabular}{lp{0.6\linewidth}}
 \toprule
  \textbf{Tool} & \textbf{Description} \\
  \midrule
  Radare2Tool & Provides Radare2 commands for binary analysis including disassembly, decompilation, and control flow analysis. \\
  \addlinespace
  GrighraTool & Provides Grighra commands for reverse engineering and analysis. \\
 \bottomrule
\end{tabular}
\end{table}

To evaluate the feasibility of {\tech}, we implemented a prototype system in Python.
The system consists of the following core components:

\begin{itemize}
\item \textit{Tools}:
Each agent is equipped with the tools summarized in Table~\ref{tab:tools}.
In particular, \textit{radare2} serves as a lightweight, scriptable reverse engineering framework that provides essential functionalities for binary analysis. It enables the agent to extract raw assembly code, cross-references (xrefs), pseudocode (when available), and relevant metadata such as strings or function information from binaries.

\item \textit{Output Schema}: Each agent follows a JSON schema (\textit{thought}, \textit{action}, \textit{action\_input}, \textit{status}). 
Here, \textit{thought} represents the agent’s reasoning or decision rationale; \textit{action} denotes the selected tool (from Table~\ref{tab:tools}); \textit{action\_input} specifies the tool parameters; and \textit{status} records the execution result or any associated error message.

\item \textit{Response Parsing}:
Structured JSON responses are extracted from the LLM’s output using regex matching, schema validation, and a bounded feedback–retry mechanism to ensure both syntactic and semantic robustness.

\item \textit{Memory}:
The conversation history is maintained as a sequence of role–content message pairs (\textit{system}, \textit{user}, \textit{assistant}, \textit{tool}, \textit{error}, and \textit{parse error}).
This memory design provides a persistent context for agent reasoning and decision-making.
\end{itemize}

All modules of {\tech} are implemented from scratch without relying on existing frameworks.
The source code is publicly available at: \url{https://github.com/bjtu-SecurityLab/FORGE}.

\section{Prompt}
\label{appendix:prompt}

This section details the system prompts used to guide the LLM-based agents for both the vulnerability discovery and validation phases. 

\subsection{System Prompt for Vulnerability Discovery}
The following prompt is used to initialize all agents during the discovery phase. 
It establishes the core principles of operation for finding new vulnerabilities.

\begin{lstlisting}[style=jsonstyle, title={System prompt for the discovery phase.}, label={lst:discovery-prompt}]
You are a professional binary security analyst. Your mission is to comprehensively analyze the specified binary file, identify all externally controllable taint sources, delegate tracing to function-level agents, and report all exploitable paths.

**Core Principles:**
- **Evidence-Based:** All analyses MUST be grounded in concrete evidence from the `r2` tool. No speculation.
- **Taint Identification:** Autonomously identify and validate all genuinely controllable external variables (e.g., HTTP params, NVRAM, IPC). The threat model assumes an attacker has network access and valid user credentials. Do not trace unexploitable taints.
- **Delegated Tracing:** You DO NOT perform deep taint tracing yourself. Your role is to delegate the tracing of specific data flows to function-level agents, ensuring a complete path from source to sink is reconstructed.
- **Focused Analysis:** Concentrate strictly on your assigned task. Your final output must be a comprehensive list of all evidence-based, exploitable paths. DO NOT provide fix suggestions or subjective commentary.
\end{lstlisting}

\subsection{System Prompt for Vulnerability Validation}
During the validation phase, agents are initialized with a specific taint propagation path and guided by the following prompt to ensure rigorous verification.

\begin{lstlisting}[style=jsonstyle, title={System prompt for the validation phase.}, label={lst:validation-prompt}]
You are a binary call chain validation agent. Your SOLE mission is to strictly verify a given call chain provided as a clue in the specified binary.

**Verification Requirements:**
- **Evidence-Only:** Base all judgments exclusively on evidence from `r2`. No guessing.
- **Path Verification:** Confirm if a reproducible propagation path exists from the specified source to the sink. Verify if the taint is genuinely exploitable under the defined threat model.
- **Success Case:** If verified, output the complete, evidence-backed propagation path with corrected addresses.
- **Failure Case:** If not verified, clearly state the reason (e.g., path does not exist, data is sanitized, sink is not reached).
\end{lstlisting}

\section{Example for Vulnerability Validation}
\label{appendix:validation_example}

The validation phase takes a structured path from the discovery phase as its input. 
This process is crucial for eliminating false positives and confirming the exploitability of a potential vulnerability. 
The following example demonstrates how the system validates the CI vulnerability discovered in our analysis.
 
\subsection{Evidence Chain}

The validation process is initiated with an evidence chain (JSON) generated during discovery.

\begin{lstlisting}[style=jsonstyle, title={A vulnerability and its Evidence Chain}, label={lst:discovery-evidence}]
{
    "type": "CWE-78",
    "identifier": ["dev_name"],
    "propagation": [
        "Source: HTTP query parameter 'dev_name' received from client request via QUERY_STRING environment variable",
        "Step: uVar1 = sym.get_querry_var(puVar3[-0x204], 0x1aef0) --> User input from 'dev_name' parameter is retrieved at 0xa730",
        "Step: sym.imp.snprintf(puVar3 + -0x808 + -4, 0x800, 0x1af04, puVar3[-3]) --> Tainted user input is formatted into command string 'cfm post netctrl 51?op=3,string_info=%s' at 0xa7ac",
        "Sink: sym.imp.system(puVar3 + -0x808 + -4) --> Tainted command string is passed to system() at 0xa7c0"
    ],
    "reason": "process_datamanage_usbeject function retrieves user-controlled dev_name parameter via get_querry_var, directly formats it into a system command using snprintf, and executes via system().",
    "file_path": "../../karonte/Tenda/../_US_AC18V1.0BR_V15.03.05.05../squashfs-root/usr/bin/app_data_center"
}
\end{lstlisting}

\subsection{Verified Evidence Chain}

After the verification, the agent produces a verified evidence chain (JSON). 
This report not only confirms the original finding but also enriches it with the missing authentication vulnerability.

\begin{lstlisting}[style=jsonstyle, title={A vulnerability and its Verified Evidence Chain.}, label={lst:validated-evidence}]
{
    "accuracy": "accurate",
    "vulnerability": true,
    "propagation": [
        "Source: An unauthenticated HTTP request is made to the path containing 'usbeject'.",
        "Step: At 0x9e6c, the dispatcher `do_request_process` matches the 'usbeject' string (located at 0x1ad18) and, without any authentication check, calls `process_datamanage_usbeject` at 0x9e8c.",
        "Step: Inside `process_datamanage_usbeject`, `sym.get_querry_var` is called at 0xa730 to retrieve the 'dev_name' string from the query parameters.",
        "Step: At 0xa7ac, the retrieved string is formatted into a command buffer using `sym.imp.snprintf` with format string 'cfm post netctrl 51?op=3,string_info=%s' (located at 0x1af04).",
        "Sink: At 0xa7c0, the unsanitized command buffer is passed directly to `sym.imp.system` for execution."
    ],
    "reason": "Verification successful. The data flow from the external source 'dev_name' to the `system` sink is confirmed. The propagation path is correct. Furthermore, analysis of the calling function `do_request_process` revealed that this entire code path is accessible without authentication, escalating the vulnerability from a standard Command Injection to an Unauthenticated Remote Code Execution. A PoC can be constructed by injecting shell metacharacters (e.g., `; reboot`) into the 'dev_name' parameter of a request to the 'usbeject' endpoint."
}
\end{lstlisting}

\section{Example of Vulnerability Discovery}
\label{appendix:foa_code_example-1}


\subsection{Turn 1: Identify the source}
The system initiates the analysis by identifying a high-risk hypothesis and delegating it.
\begin{lstlisting}[style=jsonstyle, title={Turn 1: The initial analysis task.}]
{
    "thought": "I've identified the binary 'app_data_center' which likely handles web requests. A common vulnerability class is command injection from web parameters. I will create a task-level agent to investigate all data flows from HTTP parameters to command execution sinks like `system()`.",
    "action": "AgentTool",
    "action_input": {
        "task": "Analyze binary 'app_data_center'. Trace data flows from all HTTP query parameters to command execution sinks."
    },
    "status": "continue"
}
\end{lstlisting}

\subsection{Turn 2: Root Agent -- Function-level Delegation}
\label{appendix:foa_code_example-2}

The \textbf{Root Agent} finds a potential entry point and dispatcher, then delegates deeper analysis.
\begin{lstlisting}[style=jsonstyle, title={Turn 2: Root Agent identifies the dispatcher and vulnerable function.}]
{
    "thought": "Using `r2`, I've identified `do_request_process` as a central request dispatcher. It compares the URL path to several strings. One path, 'usbeject' (at 0x1ad18), leads to a call to `process_datamanage_usbeject` at 0x9e8c. Unlike other paths, this one lacks a call to `authorization_check`. This is highly suspicious. I will delegate the analysis of `process_datamanage_usbeject` to a function-level agent, noting the lack of authentication.",
    "action": "AgentTool",
    "action_input": {
        "task": "Analyze `process_datamanage_usbeject`. Trace all external inputs within this function to potential sinks. Context: The entry to this function appears to be unauthenticated."
    },
    "status": "continue"
}
\end{lstlisting}

\subsection{Turn 3: Child Agent -- Local Discovery and Reporting}
\label{appendix:foa_code_example-3}

The \textbf{Child Agent} performs the focused analysis and finds the direct command injection chain.
\begin{lstlisting}[style=jsonstyle, title={Turn 3: Child Agent identifies the sink and prepares its local report.}]
{
    "thought": "I've analyzed `process_datamanage_usbeject`. The function retrieves the 'dev_name' parameter via `get_querry_var` at 0xa730. This value is used in an `snprintf` call at 0xa7ac, and the resulting buffer is passed directly to `system()` at 0xa7c0. This is a clear, unsanitized path from source to sink. I will report this path segment.",
    "action": "finish",
    "action_input": {
        "final_response": {
            "status": "SINK_REACHED",
            "path_segment": [
                "Source: 'dev_name' parameter retrieved via `get_querry_var` at 0xa730.",
                "Step: `snprintf` at 0xa7ac formats the value into a command.",
                "Sink: `system` at 0xa7c0 executes the command."
            ],
            "reason_snippet": "Unsanitized user input 'dev_name' flows directly into a `system` call."
        }
    },
    "status": "complete"
}
\end{lstlisting}

\subsection{Turn 4: Root Agent -- Aggregating and Relaying Upwards}
\label{appendix:foa_code_example-4}

The \textbf{Root Agent} receives the report, combines it with its own crucial context (the missing authentication), and reports the full, severe vulnerability.
\begin{lstlisting}[style=jsonstyle, title={Turn 4: Root Agent combines its context with the child's report.}]
{
    "thought": "My child agent has confirmed a command injection sink. When combined with my earlier discovery that the 'usbeject' path is unauthenticated, the full picture emerges: an unauthenticated remote command execution vulnerability. I will prepend my findings about the dispatcher to the child's path segment and report this complete path to the root agent.",
    "action": "finish",
    "action_input": {
        "final_response": {
            "status": "PATH_COMPLETE",
            "full_path": [
                "Source: Unauthenticated request to 'usbeject' endpoint handled by `do_request_process`.",
                "Step: No authentication check performed before calling `process_datamanage_usbeject` at 0x9e8c.",
                "Step: 'dev_name' parameter retrieved via `get_querry_var` at 0xa730.",
                "Step: `snprintf` at 0xa7ac formats the value into a command.",
                "Sink: `system` at 0xa7c0 executes the command."
            ],
            "reason_snippet": "A missing authentication check allows an unsanitized user parameter to reach a `system` call."
        }
    },
    "status": "complete"
}
\end{lstlisting}

\subsection{Turn 5: Final Report Assembly and Storage}
\label{appendix:foa_code_example-5}

Finally, the system receives the complete path, assembles the final, user-facing report, and assigns a critical risk score.
\begin{lstlisting}[style=jsonstyle, title={Turn 5: Root agent assembles the final, structured report for storage.}]
{
    "thought": "I have received a complete, evidence-backed vulnerability path for an Unauthenticated RCE. This is a critical finding. I will assemble the final structured report with all details from the agent chain and assign a risk score reflecting the maximum severity.",
    "action": "finish", 
    "action_input": {
        "final_response": {
            "type": "CWE-78",
            "additional_weaknesses": ["CWE-862"],
            "identifier": ["dev_name"],
            "propagation": [
                "Source: HTTP query parameter 'dev_name' on the unauthenticated '/usbeject' endpoint.",
                "Step: `do_request_process` dispatches to `process_datamanage_usbeject` without authentication.",
                "Step: At 0xa730, `get_querry_var` retrieves the user-supplied 'dev_name' string.",
                "Step: At 0xa7ac, `snprintf` formats this string into a command.",
                "Sink: At 0xa7c0, the malicious command string is executed by `system`."
            ],
            "reason": "The function `process_datamanage_usbeject` contains a command injection vulnerability via the 'dev_name' parameter. Crucially, the calling function `do_request_process` fails to perform an authentication check for the 'usbeject' endpoint, making the vulnerability accessible without credentials. This combination allows for Unauthenticated Remote Code Execution.",
            "risk_score": 9.0,
            "confidence": 9.0,
            "file_path": "../../karonte/Tenda/../_US_AC18V1.0BR_V15.03.05.05../squashfs-root/usr/bin/app_data_center"
        }
    },
    "status": "complete"
}
\end{lstlisting}
\end{document}